\documentstyle[a4p,12pt,epsfig,cite]{article}

\def\Zzero      {\mathrm{Z}^0}
\def\Kzero      {\mathrm{K}^0_s}
\def\Afb        {A_{FB}} 
\def\PL         {P_L} 
 
\def\costh      {\cos\theta^{*}} 
\def\epem       {\mathrm{e}^+\mathrm{e}^-}
\def\lbar       {\overline{\Lambda}}
\def\cosrp      {\widehat{\left(\vec{r},\vec{p}\right)}} 
 
\newcommand{\JETSET}{{\mbox{JETSET}}}
\begin{document}
\begin{titlepage}
\begin{center}{\large   EUROPEAN LABORATORY FOR PARTICLE PHYSICS}
\end{center}
\bigskip
\begin{flushright}
CERN-PPE/97-104\\
30th July 1997 
\end{flushright}
\bigskip\bigskip\bigskip\bigskip\bigskip
\begin{center}{\huge\bf Polarization and Forward-Backward}
\end{center}
\vspace{0.1cm} 
\begin{center}{\huge\bf\boldmath Asymmetry of $\Lambda$ Baryons}
\end{center}
\begin{center}{\huge\bf\boldmath in Hadronic Z$^0$ Decays}
\end{center}
\bigskip
\smallskip
\begin{center}
{\LARGE The OPAL Collaboration}
\end{center}
\bigskip
\bigskip
\begin{center}{\large  Abstract}\end{center}
\noindent 

The longitudinal polarization, the transverse polarization, and the 
forward-backward asymmetry of $\Lambda$ baryons, have been measured using a 
sample of 4.34~million hadronic $\Zzero$ decays collected with the OPAL 
detector at LEP between 1990 and 1995. These results are important as an aid 
to the understanding of hadronization mechanisms. Significant longitudinal 
polarization has been observed at intermediate and high momentum. For $x_E$ 
($\equiv 2 E_{\Lambda}/\sqrt{s}) >$~0.3, the longitudinal polarization has been 
measured to be $-32.9 \pm 5.5 $ (stat) $\pm$ 5.2 (syst)\%.  We have observed no 
transverse polarization. A significant forward-backward asymmetry has been 
measured and can be described by a JETSET model. 
\bigskip
\bigskip
\bigskip
\bigskip
\bigskip
\bigskip
\bigskip
\begin{center}
{\large (Submitted to Z. Phys. C)}
\end{center}
\vfill
\end{titlepage}
\begin{center}{\Large        The OPAL Collaboration
}\end{center}\bigskip
\begin{center}{
K.\thinspace Ackerstaff$^{  8}$,
G.\thinspace Alexander$^{ 23}$,
J.\thinspace Allison$^{ 16}$,
N.\thinspace Altekamp$^{  5}$,
K.J.\thinspace Anderson$^{  9}$,
S.\thinspace Anderson$^{ 12}$,
S.\thinspace Arcelli$^{  2}$,
S.\thinspace Asai$^{ 24}$,
D.\thinspace Axen$^{ 29}$,
G.\thinspace Azuelos$^{ 18,  a}$
% ...
}\end{center}\bigskip
\bigskip
% ...
$^{  2}$Dipartimento di Fisica dell' Universit\`a di Bologna and INFN,
I-40126 Bologna, Italy
\newline
% ...
$^{  5}$Cavendish Laboratory, Cambridge CB3 0HE, UK
\newline
% ...
$^{  8}$CERN, European Organisation for Particle Physics,
CH-1211 Geneva 23, Switzerland
\newline
$^{  9}$Enrico Fermi Institute and Department of Physics,
University of Chicago, Chicago IL 60637, USA
\newline
% ...
$^{ 12}$Indiana University, Department of Physics,
Swain Hall West 117, Bloomington IN 47405, USA
\newline
% ...
$^{ 16}$Department of Physics, Schuster Laboratory, The University,
Manchester M13 9PL, UK
\newline
% ...
$^{ 18}$Laboratoire de Physique Nucl\'eaire, Universit\'e de Montr\'eal,
Montr\'eal, Quebec H3C 3J7, Canada
\newline
% ...
$^{ 23}$Department of Physics and Astronomy, Tel Aviv University,
Tel Aviv 69978, Israel
\newline
$^{ 24}$International Centre for Elementary Particle Physics and
Department of Physics, University of Tokyo, Tokyo 113, and
Kobe University, Kobe 657, Japan
\newline
% ...
$^{ 29}$University of British Columbia, Department of Physics,
Vancouver BC V6T 1Z1, Canada
\newline
% ...
$^{  a}$ and at TRIUMF, Vancouver, Canada V6T 2A3
\newline
% ...

\newpage

\section{Introduction}

It is well known that the weak interactions are parity violating. 
Consequently, fermions produced in $\Zzero$ decays have a longitudinal 
polarization that depends on their left and right electroweak couplings. The
polarization is large for quarks but not directly measurable. Assuming the 
electroweak Standard Model to hold, the extent to which this longitudinal 
polarization is transferred to the observed hadrons is an interesting test of
hadronization. Moreover, in the absence of beam polarization, the quarks have
no transverse polarization component and any observed transverse polarization 
can only arise during the hadronization phase.  

  In the case of unpolarized beams, standard electroweak theory predicts 
for the charge $-\frac{1}{3}$ quarks from $\Zzero$ decay a longitudinal 
polarization of $-0.94$~\cite{lund_pred}. The corresponding antiquarks 
have the same degree of polarization, but with opposite helicity. The 
polarization varies by $\pm$2\% with the centre-of-mass production polar 
angle $\theta$. Gluon radiation will reduce this value slightly $-$ a one-loop 
QCD calculation~\cite{gluon_rad} shows that this reduction is about 3\%. 
Therefore, we expect that strange quarks from $\Zzero$ decays will have a 
polarization of $-91$\%, with an uncertainty of a few percent.  
 
  The possibility to measure the quark helicities from the polarization of 
the leading baryons was first suggested in \cite{renard}. In the simple quark 
model the $\Lambda$\footnote{Unless otherwise specified the use of a particle 
name refers to the particle plus the corresponding antiparticle.} spin is given
by the spin of the constituent s quark, and the polarization of the primary s 
quark will be transferred to a directly produced leading $\Lambda$. According 
to the \JETSET\ Monte Carlo program~\cite{jetset} a fraction of $\Lambda$ will 
contain the primary s quark from $\epem \rightarrow \Zzero \rightarrow 
\mathrm{s} \overline{\mathrm{s}}$, in particular those with high momentum. It 
is these $\Lambda$, through a study of their weak decays, that are thus expected
to reveal the most information about the polarization of the initial s quark. In
addition, during decays of heavier baryon resonances containing the original  
quark some of the initial quark polarization is expected to be transferred to 
the final $\Lambda$. 

  Parity violation in the weak decay of $\Lambda \rightarrow \mathrm{p}\pi$ 
leads to an angular distribution of the decay proton given by 

\begin{equation} 
\frac{1}{N}\frac{{\mathrm d}N}{{\mathrm d}\costh} = 1 + \alpha\PL\costh  
\end{equation} 

\noindent 
where $\costh$ is the angle between the proton and the decaying $\Lambda$ flight
directions, transformed to the $\Lambda$ rest frame, and $\alpha = 0.642 \pm 
0.013$ is the $\Lambda$ decay parameter~\cite{PDG}. The longitudinal 
polarization, $\PL$, of the $\Lambda$ can therefore be determined from the 
distribution of $\costh$. The dependence of $\PL$ on $\Lambda$ momentum is 
investigated by determining the polarization in a number of $x_E$ intervals, 
where the fractional energy of the $\Lambda$, $x_E$, is given by $2E_{\Lambda}/
\sqrt{s}$.  

  Several groups have reported that the $\Lambda$, as well as 
other hyperons, produced in fixed-target hadroproduction experiments exhibit 
significant transverse polarization~\cite{trans1}. However, there is no 
generally accepted mechanism that can explain the pattern of observed 
polarizations. It has been suggested~\cite{trans2} that a measurement of the 
transverse polarization of $\Lambda$ in $\epem$ annihilation events could 
indicate the extent to which final-state interactions contribute to the 
transverse polarization seen in fixed-target experiments.    

  The Standard Model of electroweak interactions predicts a forward-backward 
asymmetry of the fermions produced in $\epem$ collisions. Pairs of fermions 
are produced preferentially with the fermion forward and the antifermion 
backward with respect to the direction of the incoming electron beam. We would 
expect that those $\Lambda$ which contain the primary s quark would reflect 
this asymmetry, although again there will be a reduction due to fragmentation 
effects~\cite{light_quarks}.  

  This paper is organized as follows. The OPAL detector and event samples are
described in Section~2. In Section~3 the selection criteria for $\Lambda$
baryons are given, along with the method of determining the signal and 
background. The measurement of the longitudinal polarization and associated 
systematic error are detailed in Sections~4 and 5. In Section~6 we use the 
\JETSET\ Monte Carlo to predict the longitudinal polarization that we may  
expect to observe. Section~7 presents the measurement of the transverse 
polarization and Section~8 the measurement of the $\Lambda$ forward-backward 
asymmetry.  

\section{The OPAL Detector and Data Selection}

OPAL is a multipurpose detector covering almost the entire solid angle around 
one of four interaction regions at LEP. Details concerning the detector and its 
performance are given elsewhere~\cite{opal}. This analysis relies mainly 
on the information from the central tracking chambers which are described 
briefly in this section. 

  Tracking of charged particles is performed by a central drift chamber 
system, consisting of a vertex chamber, a jet chamber and 
$z$-chambers\footnote{In OPAL the coordinate system is defined such that the 
positive $z$-axis is along the direction of the electron beam, $r$ is the 
coordinate normal to the beam axis, and $\theta$ and $\phi$ are the polar and 
azimuthal angles.}. The central detector is positioned inside a solenoid, which 
provides a uniform magnetic field of 0.435~T. The vertex chamber 
is a precision drift chamber which covers the range $|\cos\theta|<0.95$.
The jet chamber is a large volume drift chamber 4~m long and 3.7~m in diameter 
which provides tracking in the $r$ - $\phi$ plane using up to 159 measured space
points and in $z$ by charge division along the wires.  The jet chamber also 
allows the measurement of the specific energy loss of charged particles, 
d$E$/d$x$. A d$E$/d$x$ resolution of 3.5\%~\cite{dEdx} has been obtained for 
tracks with $|\cos\theta|<0.7$, allowing particle identification over a large 
momentum range.  A precise measurement of the $z$-coordinate is provided by the 
$z$-chambers which surround the jet chamber and cover the range 
$|\cos\theta|<0.72$. The combination of these chambers leads to a momentum
resolution of $\sigma_{p_t}$/$p_t$ $\approx \sqrt{0.02^2 + (0.0015 \cdot
p_t)^2}$, $p_t$ being the transverse track momentum with respect to the beam 
direction in GeV, and where the first term represents the contribution from 
multiple Coulomb scattering.

  This analysis is based on the complete data sample collected between 1990 and 
1995 with centre-of-mass energies on or near the Z$^0$ peak. At these energies
the longitudinal polarization of the fermions produced in the $\Zzero$ decay is
effectively independent of the centre-of-mass energy.
With the requirement that all of the central tracking chambers be operational,
a total of 4.34 million hadronic Z$^0$ decays were selected using the criteria 
described in~\cite{hadsel} with an efficiency of $98.4 \pm 0.4\%$. The 
remaining background processes, such as $\epem \rightarrow {\rm \tau^+ 
\tau^-}$ and two-photon events, were estimated to be at a negligible level 
(0.1\% or less). Events where the thrust axis lay close to the beam axis were 
rejected by requiring $\vert\cos\theta_{\mathrm {thrust}}\vert < $ 0.9, where 
$\theta_{\mathrm {thrust}}$ is the polar angle of the thrust vector determined
from charged tracks. All selected events were used in the determination of the 
polarization while only events recorded at the $\Zzero$ peak (86\% of the 
data) were used in the measurement of the forward-backward asymmetry as it is
expected to vary rapidly with the centre-of-mass energy. 

  To determine the selection efficiencies for the $\Lambda$ baryon, we have 
used a sample of approximately 3~million JETSET~7.3 and 4~million 
JETSET~7.4 hadronic Z$^0$ decays that were processed through the full OPAL 
detector simulation program~\cite{gopal}. The versions of \JETSET\ have been 
tuned to agree with overall event shapes and various  single particle 
inclusive distributions and rates as measured by OPAL. Details of the 
parameters can be found in~\cite{mctune}. The two versions differ mainly in 
the tuning of the fragmentation parameters and the decay branching ratios of 
heavy flavour hadrons. The results obtained using the two JETSET samples 
separately were consistent, so the final results are based on the 
total combined Monte Carlo sample.  

\section{Selection of $\Lambda$ Candidates and Background Determination}

The following procedure is used to select $\Lambda \rightarrow \mathrm{p}\pi$
candidates. All pairs of oppositely charged tracks were checked for 
intersections in the $xy$ plane and required to pass the following set of 
conditions: 
\begin{itemize}
\item{each track must have a transverse momentum, $p_t$, in the $xy$ plane of
at least 150 MeV;} 
\item{the intersection radius in the $xy$ plane must lie between 1 and 150 cm 
from the primary event vertex;}
\item{$\Sigma\vert d_0 \vert$, the sum of the absolute values of $d_0$ ($d_0$
is defined as the distance of closest approach to the primary event vertex in 
the $xy$ plane) for each track, must be larger than 0.2 cm;}
\item{each track must have at least 40 jet chamber hits, and at least 4 
$z$-chamber hits in the barrel region ($\vert\cos\theta\vert < 0.72$), or a 
measurement of the track endpoint in the endcap region of the jet chamber. The 
cut on the number of jet chamber hits restricts the acceptance to 
$\vert\cos\theta\vert < 0.93$).}  
\item{$\cosrp$, the angle between the vector from the primary vertex to the 
intersection point and the summed momenta of the tracks, must be less than  
$0.5^{\circ}$ ($\simeq 8.7$ mrad);}
\item{neither track can have hits more than 2 cm away from the intersection 
point of the two tracks in the direction of the primary vertex;} 
\item{the decay angle of the higher momentum particle (assumed to be a proton),
$\vert\costh\vert$, must be less than 0.95. According to the Monte Carlo 
this requirement reduces the $\gamma \rightarrow \epem$ conversion background
to levels that are below 1\% of the total sample for $x_E$ between 0.05 and 
0.09 and to an insignificant amount above $x_E$ = 0.3. Any remaining $\gamma$ 
conversion events are therefore neglected in the following analysis.}   
\item{if the higher momentum track has more than 20 hits contributing to the
d$E$/d$x$ measurement, the probability\footnote{The $\chi^2$ probability is 
calculated from the difference between the measured and expected d$E$/d$x$ for 
a given particle type, in units of the d$E$/d$x$ resolution, assuming a Gaussian
distribution.} of the track being a proton is required to be more than 5\%.} 
\end{itemize}

  All track pairs passing these cuts are then refitted. Both tracks are 
constrained to originate from a common vertex in $z$ to improve the invariant 
mass resolution. The total $\chi^2$ of this fit is required to be acceptable
(less than 50). For all track pairs passing this cut the higher momentum track 
is assumed to be the proton. Track pairs with a summed momentum of at least 500 
MeV ($x_E$ = 0.027) are accepted for further analysis. The invariant mass of 
the pair, $M_{{\mathrm p}\pi}$, and the decay angle, $\costh$, are calculated. 
The distribution of the invariant mass versus the decay angle, after all of the 
above selection cuts, is shown in Figure~1a for part of the data sample.  
The $\Lambda$ signal can be clearly seen, along with a $\Kzero$ signal. 
Shown in Figure~1b is the projection onto the $M_{{\mathrm p}\pi}$ axis for 
the complete data sample, compared to the Monte Carlo. 

  The resolution of the $\Lambda$ mass, in both the data and Monte Carlo 
samples, was 2.3 $\pm$ 0.1 MeV for $x_E < 0.05$ and 3.5 $\pm$ 0.2 MeV for 
0.3 $< x_E <$ 0.4. No correction was applied to account for the difference 
between the magnetic field value used in the Monte Carlo and the value measured 
in the detector. For this reason there is a slight shift in mass visible in 
Figure~1b which has no effect in any of the following analyses. The selection 
efficiency for $\Lambda \rightarrow \mathrm{p}\pi$ decays was 25\% at $x_E$ = 
0.15, decreasing to 10\% at $x_E$ = 0.4. These efficiencies are compatible with 
those given in our most recent paper on strange baryon production~\cite{baryons}
in which different $\Lambda$ selection criteria were used.  

  The data from 0.027 $\leq x_E \leq 1.0$ were divided into nine $x_E$ 
intervals. After the cut on $\costh$ to remove the $\gamma \rightarrow \epem$ 
background, 19 bins of equal width between $-0.95 < \costh < 0.95$ were used in 
each interval. For each bin the number of $\Lambda$ candidates was determined
according to the procedure outlined in the remainder of this section. 

  Small, yet significant differences have been observed between the momentum 
spectra seen in the data and those given by the \JETSET\ Monte Carlo for 
$\Lambda$ baryons, and to a lesser extent for $\Kzero$ mesons. The Monte Carlo 
$\Lambda$ momentum distribution was weighted by an $x_E$ dependent factor in 
order to reproduce the distribution observed in the data~\cite{baryons}. These 
weights varied from 0.99 at $x_E$ = 0.05 to 0.88 at $x_E$ = 0.4. Similarly, the 
$\Kzero$ momentum distribution from the Monte Carlo was scaled by the ratio of 
the generated and observed momentum distributions~\cite{kzero}. These weights 
varied from 1.01 to 1.05.   

  Along with the $\Lambda$ signal there are residual backgrounds from two 
sources: random combinations of oppositely charged pairs of tracks, and 
$\Kzero$ decays. The amount of background from each of these sources inside a 
window around the $\Lambda$ mass ($M_{\Lambda} = 1.115684$ GeV~\cite{PDG}) was 
calculated from the Monte Carlo simulation. Different windows were used for
$x_E < 0.1$ and $x_E > 0.1$, as shown on Figure~1a. A wider window was used at 
higher momentum due to the degradation of the mass resolution as the momentum 
of the $\Lambda$ increases. Since the mass resolution also degrades as 
$\vert\costh\vert$ increases, the window increases in width as 
$\vert\costh\vert$ increases. Combining these two effects, the signal windows 
as a function of $\costh$ are given by 
$\vert M_{\mathrm{p}\pi}-M_{\Lambda}\vert <$ ($0.007 + 
\vert\costh\vert \times 0.008$ GeV) ($x_E < 0.1$) and $\vert M_{\mathrm{p}\pi} -
M_{\Lambda}\vert <$ ($0.009 + \vert\costh\vert \times 0.006$ GeV) ($x_E > 0.1$).

  In Table~1 the number of $\Lambda$ extracted in each $\costh$ bin is given,
along with the Monte Carlo efficiency and the fractions of $\Kzero$ and random 
background, for $x_E > 0.3$. The total number of $\Lambda$ candidates and the
calculated background fractions for each $x_E$ interval are given in Table~2.
The number of $\Lambda$ found in each bin are further corrected by the 
efficiency to obtain the distribution of the number of $\Lambda$ that were 
originally produced in the full data sample, as a function of $\costh$. From 
these distributions the longitudinal polarization as a function of $x_E$ was 
then determined.  

\begin{table}[t]
\begin{center}
\begin{tabular}{|r@{ -- }l|c|c|c|c|}
\hline
\multicolumn{2}{|c|}{$\costh$} & Efficiency & Random Background & $\Kzero$ 
Fraction & $\Lambda$ Signal \\  
\multicolumn{2}{|c|}{ }  & & Fraction & & \\ 
\hline
$-0.95$ & $-0.85$ & 0.043 & 0.421 & 0.005 & 197 $\pm$ 31 \\  
$-0.85$ & $-0.75$ & 0.075 & 0.246 & 0.010 & 342 $\pm$ 26 \\  
$-0.75$ & $-0.65$ & 0.084 & 0.219 & 0.014 & 344 $\pm$ 23 \\  
$-0.65$ & $-0.55$ & 0.103 & 0.189 & 0.019 & 384 $\pm$ 24 \\  
$-0.55$ & $-0.45$ & 0.112 & 0.191 & 0.023 & 408 $\pm$ 26 \\  
$-0.45$ & $-0.35$ & 0.122 & 0.158 & 0.034 & 404 $\pm$ 24 \\  
$-0.35$ & $-0.25$ & 0.132 & 0.122 & 0.056 & 458 $\pm$ 24 \\  
$-0.25$ & $-0.15$ & 0.143 & 0.141 & 0.065 & 430 $\pm$ 25 \\  
$-0.15$ & $-0.05$ & 0.140 & 0.121 & 0.073 & 494 $\pm$ 26 \\  
$-0.05$ & $\;\;\;0.05$ & 0.135 & 0.140 & 0.097 & 490 $\pm$ 26 \\  
0.05 & $\;\;\;0.15$ & 0.153 & 0.149 & 0.144 & 536 $\pm$ 28 \\  
0.15 & $\;\;\;0.25$ & 0.163 & 0.164 & 0.135 & 566 $\pm$ 28 \\  
0.25 & $\;\;\;0.35$ & 0.178 & 0.173 & 0.130 & 580 $\pm$ 30 \\  
0.35 & $\;\;\;0.45$ & 0.170 & 0.213 & 0.141 & 532 $\pm$ 30 \\  
0.45 & $\;\;\;0.55$ & 0.174 & 0.234 & 0.110 & 519 $\pm$ 30 \\  
0.55 & $\;\;\;0.65$ & 0.166 & 0.300 & 0.101 & 446 $\pm$ 31 \\  
0.65 & $\;\;\;0.75$ & 0.150 & 0.328 & 0.092 & 453 $\pm$ 32 \\  
0.75 & $\;\;\;0.85$ & 0.145 & 0.411 & 0.086 & 385 $\pm$ 35 \\  
0.85 & $\;\;\;0.95$ & 0.127 & 0.444 & 0.071 & 341 $\pm$ 31 \\  
\hline
\end{tabular}
\end{center}
\caption{The efficiency, random combination background fraction, $\Kzero$
fraction and the number of $\Lambda$ candidates extracted from the data sample,
as a function of $\costh$, for $x_E > 0.3$. The $\Lambda$ signal uncertainty 
includes contributions from both the random background and $\Kzero$ fraction 
uncertainties.} 
\end{table}

\section{Measurement of the Longitudinal Polarization}

To calculate the polarization for each $x_E$ range, the efficiency-corrected
$\costh$ distribution was fitted to a straight line of the form  

\begin{equation} 
\frac{1}{N}\frac{{\mathrm d}N}{{\mathrm d}\costh} = 1 + \alpha\PL\costh,   
\end{equation} 

\noindent
where $\alpha = 0.642 \pm 0.013$ is the $\Lambda$ decay parameter~\cite{PDG}. 
For $\lbar$, $\alpha = -0.642 \pm 0.013$ by CP invariance. However, since the
helicity of the $\overline{{\mathrm s}}$ quark is expected to be opposite that
of the s quark, the same slope is expected for $\Lambda$ and $\lbar$. The 
polarization is thus calculated directly from the slope of the fitted line.  

  The measured values of $\PL$, as calculated from the fits, are given in 
Table~2. From these results we observe an indication of longitudinal 
polarization for $\Lambda$ with $x_E > 0.09$ and a significant polarization 
for those with $x_E > 0.2$. The systematic errors are discussed in the 
following section, and the results are compared to a \JETSET\ Monte Carlo 
calculation in Section~6. The fits are good in all $x_E$ intervals. In Figure~2
the fits for 3 separate $x_E$ regions are shown, along with the fit for $x_E > 
0.3$, where the $\chi^2$ is 19.5 for 19 fitted points.  

  Since the $\Lambda$ has a non-zero magnetic moment the spin will precess in 
the magnetic field of the detector. We have estimated that this effect will 
cause a change in the polarization of no more than 1 -- 2\%, depending on the 
momentum of the $\Lambda$ and the polar angle of the flight direction. We have 
not included any correction for this effect in our results.

\begin{table}
\begin{center}
\begin{tabular}{|r@{ -- }l|c|c|c|r@{ $\pm$ }c@{ $\pm$ }l|}
\hline
\multicolumn{2}{|c|}{$x_E$} & Total $\Lambda$ & Random Background & 
$\Kzero$ Fraction & \multicolumn{3}{c|}{$\PL$ (\%)} \\  
\multicolumn{2}{|c|}{ } & & Fraction & & \multicolumn{3}{|c|}{ } \\ 
\hline
0.027 & 0.05 & 22280 & 0.170 & 0.092 & 1.1 & 3.8 & 3.0 \\  
0.05 & 0.08 & 42479 & 0.181 & 0.091 & $-2.5$ & 2.2 & 2.5 \\  
0.08 & 0.09 & 11542 & 0.135 & 0.036 & 0.4 & 3.9 & 2.2 \\  
0.09 & 0.1 & 10031 & 0.140 & 0.029 & $-8.9$ & 4.2 & 3.2 \\  
0.1 & 0.15 & 35101 & 0.148 & 0.034 & $-5.7$ & 2.2 & 1.3 \\  
0.15 & 0.2 & 18139 & 0.187 & 0.039 & $-9.1$ & 3.2 & 3.5 \\  
0.2 & 0.3 & 15723 & 0.218 & 0.049 & $-15.4$ & 3.7 & 3.9 \\  
0.3 & 0.4 & 5569 & 0.205 & 0.072 & $-19.3$ & 6.5 & 6.5  \\  
0.4 & 1.0 & 2950 & 0.222 & 0.110 & $-45.7$ & 9.8 & 7.7  \\  
\hline 
\hline 
0.3 & 1.0 & 8309 & 0.210 & 0.084 & $-32.9$ & 5.5 & 5.2 \\  
\hline
\end{tabular}
\end{center}
\caption{The number of $\Lambda$ candidates extracted from the data 
sample, the random background fraction, $\Kzero$ fraction, and the measured 
longitudinal polarization of $\Lambda$ from $\Zzero$ decay. The first error 
quoted for the polarization is statistical, the second systematic.}  
\end{table}

\section{Systematic Error on the Longitudinal Polarization}

Several possible sources of systematic error were identified and 
studied: background determination, the $\Lambda$ selection cuts, the acceptance 
window used to determine the number of $\Lambda$ events, and the factors used 
to renormalize the Monte Carlo $\Kzero$ and $\Lambda$ momentum distributions. 
Each of these will be discussed below and the values summarized in Tables~3 
and 4.  

  As was mentioned in the previous section, the polarization is expected to be 
the same for $\Lambda$ and $\lbar$. This was verified explicitly. For $x_E > 
0.3$ the polarization for $\Lambda$ ($\lbar$) was measured to be $-33.4 \pm$ 
7.8\% ($-32.3 \pm$ 7.9\%), where the errors are statistical only. It was also 
verified that the polarizations of $\Lambda$ and $\lbar$ were consistent in 
all momentum regions. 

  The fraction of random background in the signal is not constant as a function
of $\costh$ and increases as $\vert\costh\vert \rightarrow 1$ (see Table~1). 
If the background is not properly corrected for, especially at larger values of 
$\vert\costh\vert$, a significant systematic effect could be introduced. To 
estimate the effect of the background determination the efficiency corrected 
$\costh$ distributions were fitted over 3 different sub-intervals: $-0.75 < 
\costh < 0.95$, $-0.75 < \costh < 0.65$ and $-0.95 < \costh < 0.65$. The 
systematic error was taken to be the RMS deviation of these fitted values from 
the value obtained when fitting over the full interval.

  The selection cuts used to isolate the $\Lambda$ signal are also a possible
source of systematic error. If a particular cut variable is not well modelled by
the Monte Carlo too many or too few events will be removed in either the data
or Monte Carlo and a possible systematic error introduced. The calculation of 
the polarization was repeated for several different values of the most important
selection cuts (the $\cosrp$, $\Sigma\vert d_0 \vert$, d$E$/d$x$, and hit radius
cuts).  The cut on $\cosrp$ was varied between $0.25^{\circ}$ and $1.0^{\circ}$
and the $\Sigma\vert d_0 \vert$ cut between 0.2 and 0.4~cm. The d$E$/d$x$ 
probability was varied between 0 (no cut applied) and 10\%. The cut on the 
distance of hits away from the intersection point of the two tracks was varied 
between 1 and 10~cm. For each cut at least 6 different values of the cut were 
chosen within the ranges specified and the polarization recalculated for each 
value. The RMS deviation of these values from the value obtained using the cut 
default was taken to be the systematic error for that particular selection cut.
The contributions from each selection cut as a function of $x_E$ are given in 
Table~3, where the total systematic error due to the $\Lambda$ selection 
procedure is then obtained by adding the contributions from all cuts in 
quadrature.   

\begin{table}
\begin{center}
\begin{tabular}{|r@{ -- }l|c|c|c|c||c|}
\hline 
\multicolumn{2}{|c|}{ } & \multicolumn{5}{|c|}{Selection Cut} \\ 
\cline{3-7} 
\multicolumn{2}{|c|}{$x_E$} & $\cosrp$ & $\Sigma\vert d_0 \vert$ & Hit Radius 
& d$E$/d$x$ & Total \\
\hline
0.027 & 0.05 & 1.9 & 0.4 & 1.1 & 0.7 & 2.3 \\   
0.05 & 0.08 & 1.8 & 0.6 & 0.6 & 0.9 & 2.2 \\   
0.08 & 0.09 & 1.5 & 0.8 & 0.3 & 0.5 & 1.8 \\   
0.09 & 0.1 & 1.0 & 0.3 & 0.7 & 0.6 & 1.9 \\   
0.1 & 0.15 & 0.7 & 0.4 & 0.5 & 0.6 & 1.1 \\   
0.15 & 0.2 & 1.7 & 0.7 & 0.5 & 1.0 & 2.2 \\   
0.2 & 0.3 & 0.8 & 1.7 & 0.4 & 1.5 & 2.4 \\   
0.3 & 0.4 & 1.4 & 2.2 & 0.9 & 1.1 & 3.0 \\
0.4 & 1.0 & 2.5 & 3.4 & 2.0 & 3.2 & 5.6 \\   
\hline 
\hline 
0.3 & 1.0 & 1.6 & 2.2 & 1.0 & 1.1 & 3.1 \\   
\hline
\end{tabular}
\end{center}
\caption{Systematic error contributions (in \%) to $\PL$ from the selection 
cuts. The total contribution is the sum, added in quadrature, of the individual 
contributions.} 
\end{table}

  The width of the acceptance window around the $\Lambda$ mass may also affect 
the calculation of $\PL$. If the window is too narrow, $\Lambda$ that are 
reconstructed with an invariant mass a significant distance away from the 
nominal $\Lambda$ mass will be lost which could introduce a $\costh$ bias. 
Since the resolution in the data is almost the same as in the Monte Carlo, it 
is expected that as long as the window is wide enough to include the signal 
peak the systematic effects will be small. To investigate the effect of the 
acceptance window, several additional windows were studied, both narrower and 
wider than the default. As in the two previous cases, the RMS deviation of the 
values from the default case was taken to be the systematic error. 

  The method used to adjust the Monte Carlo momentum distributions such that
they agreed with those observed experimentally is another possible source of 
systematic error. Weighting factors were determined as a function of momentum
and applied to the Monte Carlo distributions. These factors each had an 
experimental error associated with them. The factors were varied by $\pm 
1\sigma$  and the polarization recalculated. The difference from the 
value determined using the nominal correction factor was taken as the 
systematic error.  

  The various contributions to the systematic error are compiled in Table~4. 
The total systematic error is obtained by adding all of the contributions in 
quadrature. 

\begin{table} 
\begin{center}
\begin{tabular}{|r@{ -- }l|c|c|c|c||c|}
\hline 
\multicolumn{2}{|c|}{$x_E$} & Background & Selection & Acceptance & Momentum
  & Total \\
\multicolumn{2}{|c|}{ } & Determination & Cuts & Window & Distribution & \\  
\hline
0.027 & 0.05 & 0.7 & 2.3 & 1.8 & 0.3 & 3.0 \\   
0.05 & 0.08 & 0.1 & 2.2 & 1.2 & 0.2 & 2.5 \\   
0.08 & 0.09 & 1.1 & 1.8 & 0.4 & 0.4 & 2.2 \\   
0.09 & 0.1 & 2.5 & 1.9 & 0.5 & 0.5 & 3.2 \\   
0.1 & 0.15 & 0.4 & 1.1 & 0.4 & 0.4 & 1.3 \\   
0.15 & 0.2 & 2.3 & 2.2 & 1.2 & 0.6 & 3.5 \\   
0.2 & 0.3 & 2.9 & 2.4 & 0.7 & 0.7 & 3.9 \\   
0.3 & 0.4 & 5.4 & 3.0 & 1.7 & 1.0 & 6.5 \\   
0.4 & 1.0 & 4.4 & 5.6 & 2.7 & 1.1 & 7.7 \\   
\hline 
\hline 
0.3 & 1.0 & 3.2 & 3.1 & 2.3 & 1.2 & 5.2 \\   
\hline
\end{tabular}
\end{center}
\caption{Contributions (in \%) to the systematic error on $\PL$.}  
\end{table}

\section{Longitudinal Polarization Prediction}

We use the model of Gustafson and H\"{a}kkinen~\cite{lund_pred} to calculate 
the expected polarization from each of several $\Lambda$ production sources. 
The \JETSET\ Monte Carlo has been used to determine the production rates of 
$\Lambda$ from each of these sources, and the calculated polarizations are 
then compared to the measurements. 

  It is necessary to distinguish between $\Lambda$ which are produced directly
and those which come from decays of heavier resonances. It is also necessary to
distinguish between $\Lambda$ which contain a primary strange quark and those
which contain an s quark produced in the fragmentation process. To estimate 
the polarization in each of these cases, the following assumptions are 
made~\cite{lund_pred}:
\begin{itemize} 
\item{In a simple quark model the spin of the $\Lambda$ is determined by the 
spin of the s quark. Therefore, a directly produced $\Lambda$ should be 
polarized in the same way as the primary s quark.}  
\item{Some $\Lambda$ will be decay products of heavier baryons which contain
the primary s quark. These $\Lambda$ will inherit some fraction of the parent's
polarization. Estimates are given for the $\Sigma^0$, $\Xi^-$, 
$\Sigma(1385)^{\pm}$, and $\Xi(1530)$.}  
\item{If a $\Lambda$ contains a primary u or d quark, that quark becomes part
of a spin-0 ud diquark pair, and it is assumed the polarization of the initial
quark is lost in the formation of the $\Lambda$.} 
\item{Quarks produced in the fragmentation process are expected to have no 
longitudinal polarization and consequently $\Lambda$ containing these quarks
will not be polarized.} 
\end{itemize}

  It is therefore expected that much of the contribution to an observable 
polarization will come from $\Lambda$ that contain the initial s quark, either
directly or via a decay. Contributions from the decay of the $\Omega^-$ can be 
neglected due to its very low production rate. There will likely be some 
contribution from the decays of charm and beauty baryons, but just how much 
of the quark polarization that will be transferred to the $\Lambda$ is unknown.

  It is possible to use the \JETSET\ Monte Carlo to predict the relative 
abundances of $\Lambda$ from the various production sources as a function of 
$x_E$. Two versions of \JETSET\ version 7.4 were used. The first version uses 
the default version of the ``popcorn" model for baryon production~\cite{popcorn}
and has been tuned by OPAL. The parameter set used (see~\cite{mctune}) results 
from a global fit to OPAL measurements of event shape distributions, mean 
charged particle multiplicities, single particle inclusive momentum spectra 
for $\pi^{\pm}$, ${\mathrm K}^{\pm}$, p($\overline{\mathrm p}$) and $\Lambda$
($\lbar$); and to LEP measurements of the single particle inclusive production 
rates of 26 hadrons identified in $\Zzero$ decays. Even with this additional 
tuning significant differences still exist between the predicted and observed 
rates and momentum spectra of several baryons~\cite{baryons}. Recently however, 
a modified popcorn scheme has been proposed~\cite{new_popcorn} which 
incorporates a more complete implementation of baryon production within the 
LUND string fragmentation model.  This version is also used with the OPAL 
tuning, together with an adjustment of PARJ(10)~\footnote{In the modified 
popcorn scheme there is a suppression of leading baryons. In order to obtain
the proper amount of final baryons, a scale factor (PARJ(10)) is applied to
PARJ(1), the suppression factor for diquark production.} to 2.8, to obtain 
better agreement with the observed baryon rates and the $\Lambda$ momentum 
distribution. 

  In Table 5 the sources of origin of the $\Lambda$ are given as predicted by
both the default and modified versions of the popcorn scheme for $x_E >$ 0.15,
0.3 and 0.4. The contribution to the polarization from each source is also 
given (according to the method of~\cite{lund_pred}, except that we have assumed 
an additional contribution of 25\% for $\Lambda$ from the decays of charm and 
bottom flavoured baryons). This assumption is based on a measurement of the 
$\Lambda_{\mathrm b}$ polarization at LEP~\cite{lambdab_pol}. The calculated 
polarization is shown in Figure~3 for both the default and modified versions of 
the popcorn scheme. The modified popcorn scheme predicts less polarization,
due mostly to the reduced rate of direct $\Lambda$ production in that model 
(for example, for $x_E >$ 0.3 only 13.7\% of $\Lambda$ contain the primary s 
quark directly, compared to 27.8\% in the default popcorn version). 

\begin{table}
\begin{center}
\begin{tabular}{|c|c|c|c|c|c|c|c|}
\hline 
 & & \multicolumn{2}{|c|}{$x_E > 0.15$} & \multicolumn{2}{|c|}{$x_E > 0.3$} &
\multicolumn{2}{|c|}{$x_E > 0.4$} \\ 
\cline{3-8} 
 $\Lambda$ Source & $\Lambda$ Polarization & Default & MOPS & Default & MOPS & 
Default & MOPS \\  
\hline
Fragmentation & 0. & 0.461 & 0.566 & 0.254 & 0.398 & 0.177 & 0.344 \\ 
 u & 0. & 0.020 & 0.028 & 0.036 & 0.054 & 0.044 & 0.069 \\ 
 d & 0. & 0.015 & 0.023 & 0.027 & 0.044 & 0.034 & 0.056 \\  
direct s & $-0.91$ & 0.152 & 0.071 & 0.278 & 0.137 & 0.355 & 0.175 \\ 
$\Sigma^0$ & $-0.10$ & 0.010 & 0.024 & 0.018 & 0.046 & 0.022 & 0.059 \\ 
$\Sigma^{\ast}$ & $-0.51$ & 0.055 & 0.037 & 0.096 & 0.068 & 0.110 & 0.077 \\ 
$\Xi$ & $-0.55$ & 0.058 & 0.034 & 0.103 & 0.064 & 0.123 & 0.074 \\ 
$\Xi^{\ast}$ & $-0.46$ & 0.008 & 0.011 & 0.014 & 0.020 & 0.016 & 0.022 \\ 
$\Omega^-$ & 0. & 0.001 & 0.001 & 0.001 & 0.002 & 0.001 & 0.002 \\ 
c baryon & $-0.25$ & 0.086 & 0.072 & 0.112 & 0.098 & 0.088 & 0.083 \\ 
b baryon & $-0.25$ & 0.073 & 0.063 & 0.045 & 0.046 & 0.025 & 0.031 \\ 
b meson & 0. & 0.061 & 0.070 & 0.016 & 0.023 & 0.005 & 0.008 \\  
\hline
\multicolumn{2}{|c|}{Predicted Polarization (\%)} & $-24.3$ & $-14.5$ & $-40.6$
 & $-24.4$ & $-48.4$ & $-28.4$ \\ 
\hline
\end{tabular}
\end{center}
\caption{Relative contributions to the $\Lambda$ rate for $x_E >$ 0.15, 0.3 
and 0.4 in the default version of the \JETSET\ popcorn scheme and the modified 
popcorn scheme (MOPS).}
\end{table}

  The agreement between the predicted longitudinal polarization of the two
\JETSET\ versions and the measurements is quite good over the entire momentum 
range. No clear discrimination between the two versions is possible. 
There are probably other effects that must be considered beyond the model used 
here. However, even with these simple assumptions, the observed polarization 
is reasonably well modelled.

\section{Measurement of the Transverse Polarization}

It has been observed that in hadron-hadron collisions $\Lambda$ baryons
obtain a significant polarization in the direction perpendicular to the event 
plane. Several models (see~\cite{trans1} and references therein) have been put 
forward to explain this effect but, as yet, there is no accepted explanation. 
In $\epem$ annihilation the transverse polarization of the primary quarks is 
suppressed by a factor $m_{\mathrm q}/\sqrt{s}$ from helicity conservation, so 
any transverse polarization will arise only in the hadronization 
phase~\cite{trans2}. 

  The transverse polarization is investigated along a direction, $\hat{a} =  
\hat{p}_{\Lambda} \times \hat{p}_{\mathrm {thrust}}$, where $\hat{p}_{\Lambda}$ 
is the $\Lambda$ direction and $\hat{p}_{\mathrm {thrust}}$ is the direction of 
the thrust axis in the $\Lambda$ hemisphere. The method to determine the 
transverse polarization is very similar to that used to measure the longitudinal
polarization. The function given by Equation (1) is fitted to an efficiency 
corrected distribution of $\cos\phi_{p}$, where $\phi_{p}$ is the angle in the 
$\Lambda$ rest frame between the proton and $\hat{a}$. 

  A p$\pi$ mass selection window of constant width as a function of $\cos\phi_p$
was used. The width of the window increased with $x_E$ to account for the 
worsening of the mass resolution with increasing momentum. For $x_E <$ 0.08 a 
window of width $\pm$ 8 MeV from the nominal $\Lambda$ mass was used, for $0.08
< x_E <$ 0.1 the width was $\pm$ 11 MeV, for $0.1 < x_E < 0.3$ it was $\pm$ 15
MeV and for $x_E > 0.3$ it was $\pm$ 20 MeV. All other selection cuts were the 
same as those used in the $\PL$ analysis, and the method of determining the 
efficiency-corrected distributions was identical.  

  The data were binned in $p_T$, where $p_T$ is the transverse momentum of the 
$\Lambda$ measured relative to the thrust axis. The results of the fits for 
several $p_T$ intervals are shown in Figure~4. As in the longitudinal 
polarization analysis, the transverse polarization, $P_T^{\Lambda}$, is given 
by the slope of the fitted line divided by the $\Lambda$ decay parameter, 
$\alpha$. The results of these fits are given in Table~6. Since the random 
background in this case has no significant dependence on $\cos\phi_p$, it is
assumed that there is no systematic effect due to the determination of that 
background. All other systematic error contributions were determined following  
the same procedure as was used in the $P_L$ analysis. Additionally, there were 
no significant differences found between the values measured for $\Lambda$ and 
$\lbar$. For example, for $p_T >$ 0.3 GeV/$c$, the transverse polarization 
measured for $\Lambda$ was 1.9 $\pm$ 1.4\% and for $\lbar$ the value 
obtained was 1.5 $\pm$ 1.4\% (statistical errors only).

\begin{table}
\begin{center}
\begin{tabular}{|c|r@{ $\pm$ }c@{ $\pm$ }l|}
\hline
 $p_T$ (GeV/$c$) & \multicolumn{3}{|c|}{$P_T^{\Lambda}$ (\%)} \\
\hline
$<$ 0.3 & $-1.8$ & 3.1 & 1.0 \\  
0.3 -- 0.6 & 0.4 & 1.8 & 0.7 \\  
0.6 -- 0.9 & 1.0 & 1.9 & 0.7 \\  
0.9 -- 1.2 & 0.8 & 2.2 & 0.6 \\  
1.2 -- 1.5 & 0.0 & 2.7 & 0.6 \\  
$>$ 1.5 & 1.8 & 1.6 & 0.5 \\  
\hline 
\hline 
$>$ 0.3 & 0.9 & 0.9 & 0.3 \\  
$>$ 0.6 & 1.1 & 1.0 & 0.4 \\  
\hline
\end{tabular}
\end{center}
\caption{Measured transverse polarization of $\Lambda$ baryons as a function 
of $p_T$ (the transverse momentum of the $\Lambda$ measured relative to the
event thrust axis). The first error is statistical, the second systematic.}
\end{table}

  From the results shown in Table~6 we conclude that there is no evidence for 
any significant transverse polarization of $\Lambda$ baryons over the entire 
range of $p_T$. As was mentioned previously, it is expected that the primary 
quarks will not be transversely polarized. This is investigated by applying an 
energy cut and studying only those $\Lambda$ with $x_E > 0.15$. The result for 
the transverse polarization of these $\Lambda$ is $P_T^{\Lambda} = -0.4 \pm 
2.3$\% (statistical error only) for $p_T > 0.3$~GeV/$c$. In addition, we have 
studied the transverse polarization in and out of the scattering plane. If the 
thrust axis is replaced by the $z$-axis, the transverse polarization out of 
the scattering plane, along the direction $\hat{p}_{\Lambda} \times \hat{z}$, 
is found to be $-1.1 \pm$ 1.8\%. In the scattering plane, along the direction 
defined by $\hat{p}_{\Lambda} \times (\hat{z} \times \hat{p}_{\Lambda})$, the 
polarization is found to be $-1.3 \pm$ 1.7\%. The errors are statistical 
errors only.  

\section{Measurement of the Forward-Backward Asymmetry}

The $\Lambda$ forward-backward asymmetry has also been measured over the 
same $x_E$ range as the longitudinal polarization. It is expected that for 
high momentum $\Lambda$ this asymmetry will reflect the original s quark 
asymmetry~\cite{light_quarks}. In this part of the analysis the corrected 
distributions of $B\cos\theta$ were used to determine $\Afb$, where $B$ is the 
baryon number and $\theta$ is measured with respect to the direction of the 
incoming electron beam. 

  From the efficiency-corrected distributions of $B\cos\theta$ the 
forward-backward asymmetry is calculated from: 

\begin{equation}
\Afb = \frac{N_F - N_B}{N_F + N_B} 
\end{equation} 

\noindent
where $N_F$ is the total number with $B\cos\theta$ $>$ 0 and $N_B$ the total
number with $B\cos\theta$ $<$ 0. The values of $\Afb$ calculated with Equation
(3) are given in Table~7, as well as the values calculated from the 
OPAL tuned version of the \JETSET\ Monte Carlo, with and without the modified 
popcorn scheme. The first error is statistical,
the second systematic. The random background was symmetric about $\cos\theta 
= 0$ and it was found to give a negligible contribution to the
systematic error. The other contributions to the 
systematic error were determined using the same procedures as those used  
for the longitudinal polarization analysis. In Figure~5 the measurements are 
plotted along with curves showing the predictions from both the default OPAL
version of \JETSET\ and the modified popcorn version. The agreement between 
the measurements and the \JETSET\ models is good, but the present statistics
do not allow a distinction between models. 
 
\begin{table}
\begin{center}
\begin{tabular}{|r@{ }l|r@{ $\pm$ }c@{ $\pm$ }l|c|c|} 
\hline
\multicolumn{2}{|c|}{$x_E$} & \multicolumn{3}{|c|}{$\Afb$} & $\Afb$ & $\Afb$ \\
\multicolumn{2}{|c|}{ } & \multicolumn{3}{|c|}{(measured)} & (JETSET default)
 & (modified popcorn) \\
\hline
0.027 & -- 0.05 & $-0.007$ & 0.008 & 0.006 & 0.003 & 0.003 \\  
0.05 & -- 0.08 & 0.005 & 0.005 & 0.005 & 0.005 & 0.004 \\  
0.08 & -- 0.09 & $-0.008$ & 0.009 & 0.005 & 0.009 & 0.006 \\  
0.09 & -- 0.1 & 0.022 & 0.010 & 0.006 & 0.010 & 0.008 \\  
0.1 & -- 0.15 & 0.030 & 0.005 & 0.006 & 0.018 & 0.013 \\  
0.15 & -- 0.2 & 0.033 & 0.008 & 0.006 & 0.027 & 0.022 \\  
0.2 & -- 0.3 & 0.029 & 0.008 & 0.006 & 0.040 & 0.034 \\  
0.3 & -- 0.4 & 0.068 & 0.015 & 0.008 & 0.056 & 0.043 \\  
0.4 & -- 1.0 & 0.102 & 0.021 & 0.008 & 0.074 & 0.058 \\  
\hline
\hline 
0.15 & - 1.0 & 0.047 & 0.005 & 0.006  & 0.045 & 0.036 \\ 
0.3 & - 1.0 & 0.083 & 0.012 & 0.006  & 0.062 & 0.050 \\ 
\hline
\end{tabular}
\end{center}
\caption{Experimentally determined $\Afb$ for $\Lambda$ baryons. The first error
is statistical, the second is systematic.}
\end{table}

\section{Discussion and Conclusions}

We have observed significant values of longitudinal polarization for $\Lambda$
with intermediate and high momentum. For $x_E > 0.3$ the polarization has been
measured to be 
\[ -32.9 \pm 7.6\% \] 

\noindent 
The total error is given by the statistical and systematic errors added in 
quadrature. This value is in agreement with that reported by the ALEPH
Collaboration~\cite{ALEPH} who have measured $\PL = -32 \pm 7$\% for $z = 
p/p_{\mathrm{beam}} (\simeq x_E) > 0.3$. The longitudinal polarization results
are reasonably well modelled using a simple quark model and the \JETSET\ Monte
Carlo, which has been tuned using LEP data. However, as is discussed  
in~\cite{lampol_th}, the interpretation of the results is not unique.  We have 
also investigated the transverse polarization of $\Lambda$ baryons. No 
significant evidence was found for any transverse polarization. This is 
consistent with the result reported by ALEPH~\cite{ALEPH}.

 For $x_E > 0.15$ the $\Lambda$ forward-backward asymmetry was found to be 
\[ 0.047 \pm 0.008. \]  
\noindent
This is in agreement with the value of 0.0450 $\pm$ 0.0053 reported by 
ALEPH~\cite{ALEPH}. At higher momenta, $x_E >$ 0.3, the asymmetry was measured 
to be 
\[ 0.083 \pm 0.013, \]
\noindent
again in agreement with the ALEPH result of 0.085 $\pm$ 0.012, and the
DELPHI~\cite{DELPHI} result of 0.085 $\pm$ 0.039 measured in the range $0.25
< z < 0.5$ ($z = p/p_{\mathrm{beam}}$). The measurements are also in agreement
with the expectation from JETSET. 
 
\bigskip
\bigskip
\noindent 
{\bf Acknowledgements} 
\par
We particularly wish to thank the SL Division for the efficient operation
of the LEP accelerator at all energies
 and for
their continuing close cooperation with
our experimental group.  We thank our colleagues from CEA, DAPNIA/SPP,
CE-Saclay for their efforts over the years on the time-of-flight and trigger
systems which we continue to use.  In addition to the support staff at our own
institutions we are pleased to acknowledge the  \\
Department of Energy, USA, \\
National Science Foundation, USA, \\
Particle Physics and Astronomy Research Council, UK, \\
Natural Sciences and Engineering Research Council, Canada, \\
Israel Science Foundation, administered by the Israel
Academy of Science and Humanities, \\
Minerva Gesellschaft, \\
Benoziyo Center for High Energy Physics,\\
Japanese Ministry of Education, Science and Culture (the
Monbusho) and a grant under the Monbusho International
Science Research Program,\\
German Israeli Bi-national Science Foundation (GIF), \\
Bundesministerium f\"ur Bildung, Wissenschaft,
Forschung und Technologie, Germany, \\
National Research Council of Canada, \\
Hungarian Foundation for Scientific Research, OTKA T-016660, 
T023793 and OTKA F-023259.\\

\vfill
\newpage 
\begin{figure}[htbp]
\begin{center} 
\mbox{\epsfig{file=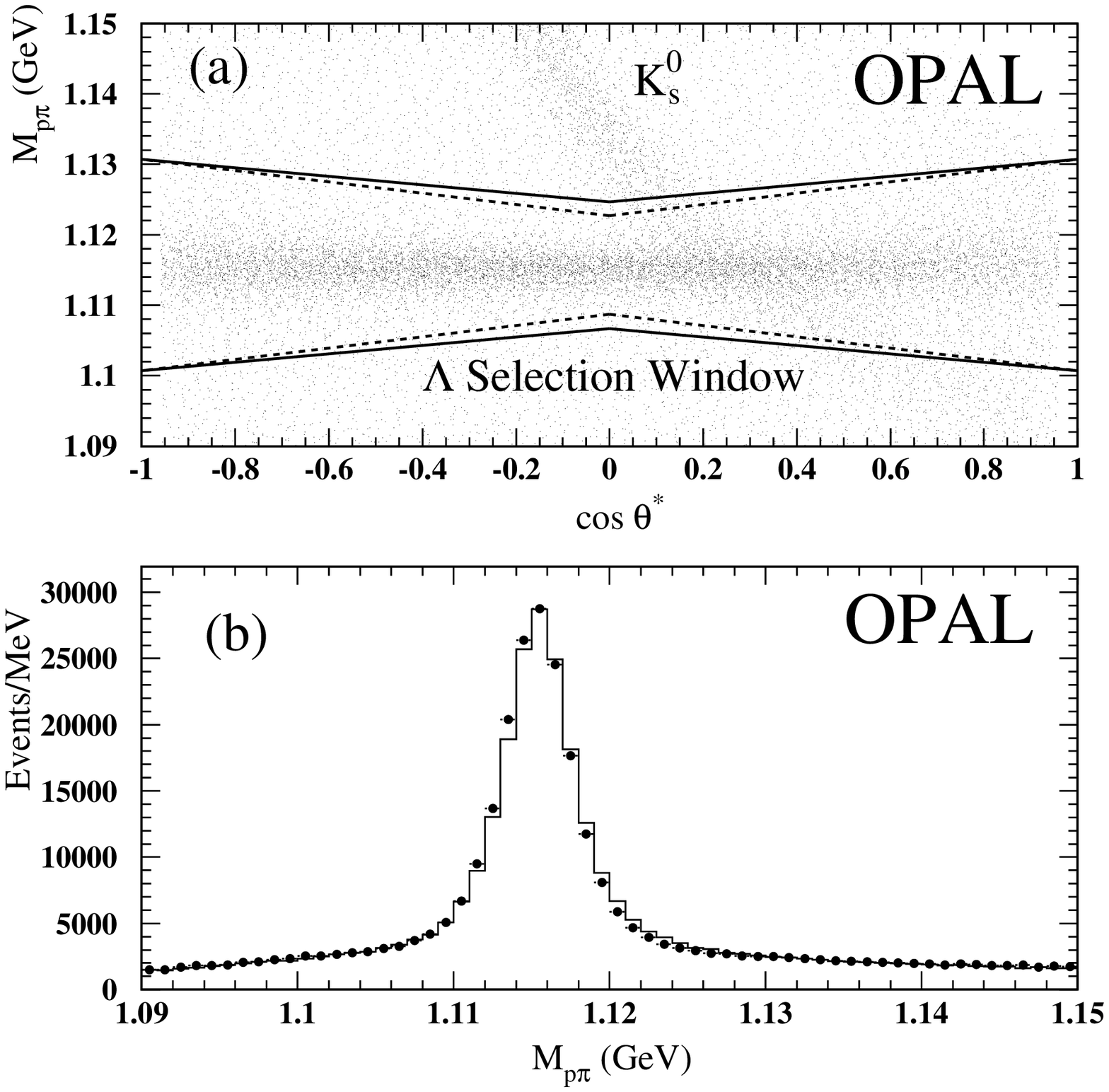,height=20cm,width=17.5cm}}
\end{center} 
\caption{(a) Invariant mass of $\Lambda$ candidates versus $\costh$ after 
selection cuts. The window for selecting $\Lambda$ is shown as a dashed line
($x_E < 0.1$) and a solid line ($x_E > 0.1$). The $\Kzero - \Lambda$ overlap
region can also be seen around $\costh = 0.2$. (b) The invariant mass of all
$\Lambda$ candidates after the selection cuts. The histogram is from the Monte 
Carlo, normalized to the number of OPAL data events (solid points).} 
\end{figure} 

\newpage 
\begin{figure}[htbp]
\begin{center} 
\mbox{\epsfig{file=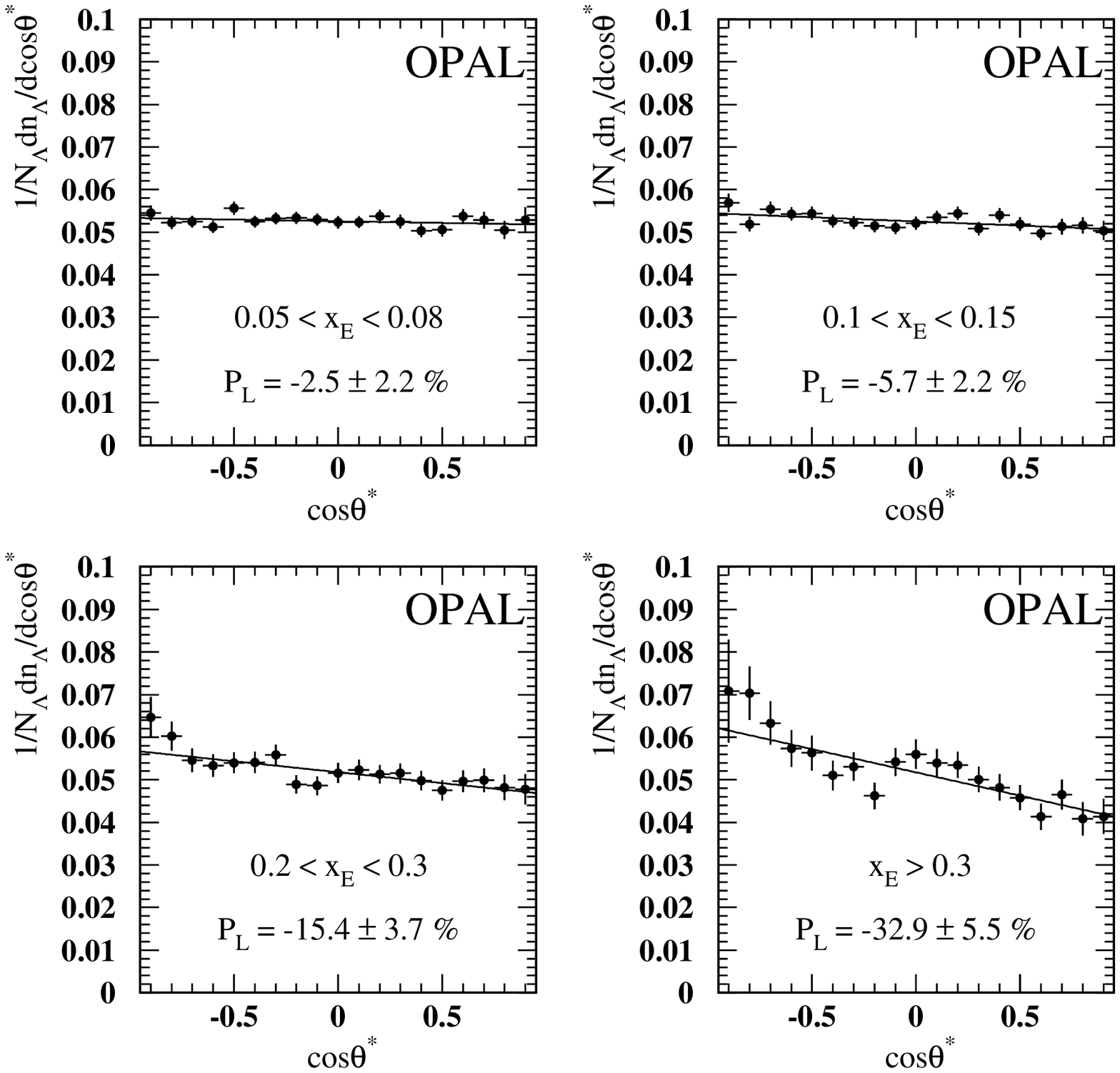,height=20cm,width=17.5cm}}
\end{center} 
\caption{The fits to the efficiency-corrected $\costh$ distributions for 
several $x_E$ regions. The error bars are statistical errors only.} 
\end{figure} 

\newpage 
\begin{figure}[htbp]
\begin{center} 
\mbox{\epsfig{file=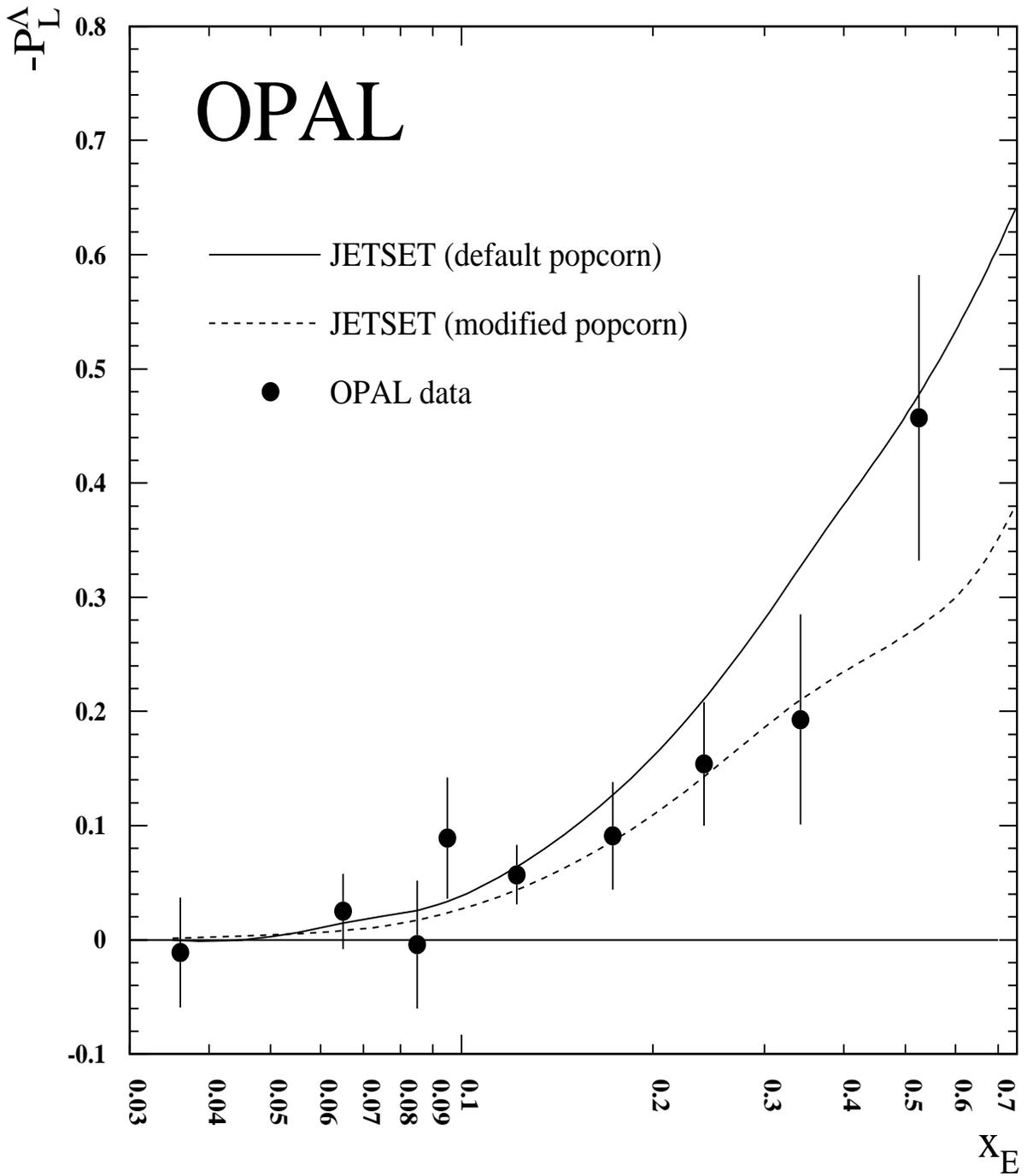,height=20cm,width=17.5cm}}
\end{center} 
\caption{The longitudinal polarization predicted using tuned versions of
\JETSET\ with different versions of the popcorn model of baryon production. 
The measurements are shown as solid points and the error bars are the 
statistical plus systematic errors, added in quadrature.} 
\end{figure} 

\newpage 
\begin{figure}[htbp]
\begin{center} 
\mbox{\epsfig{file=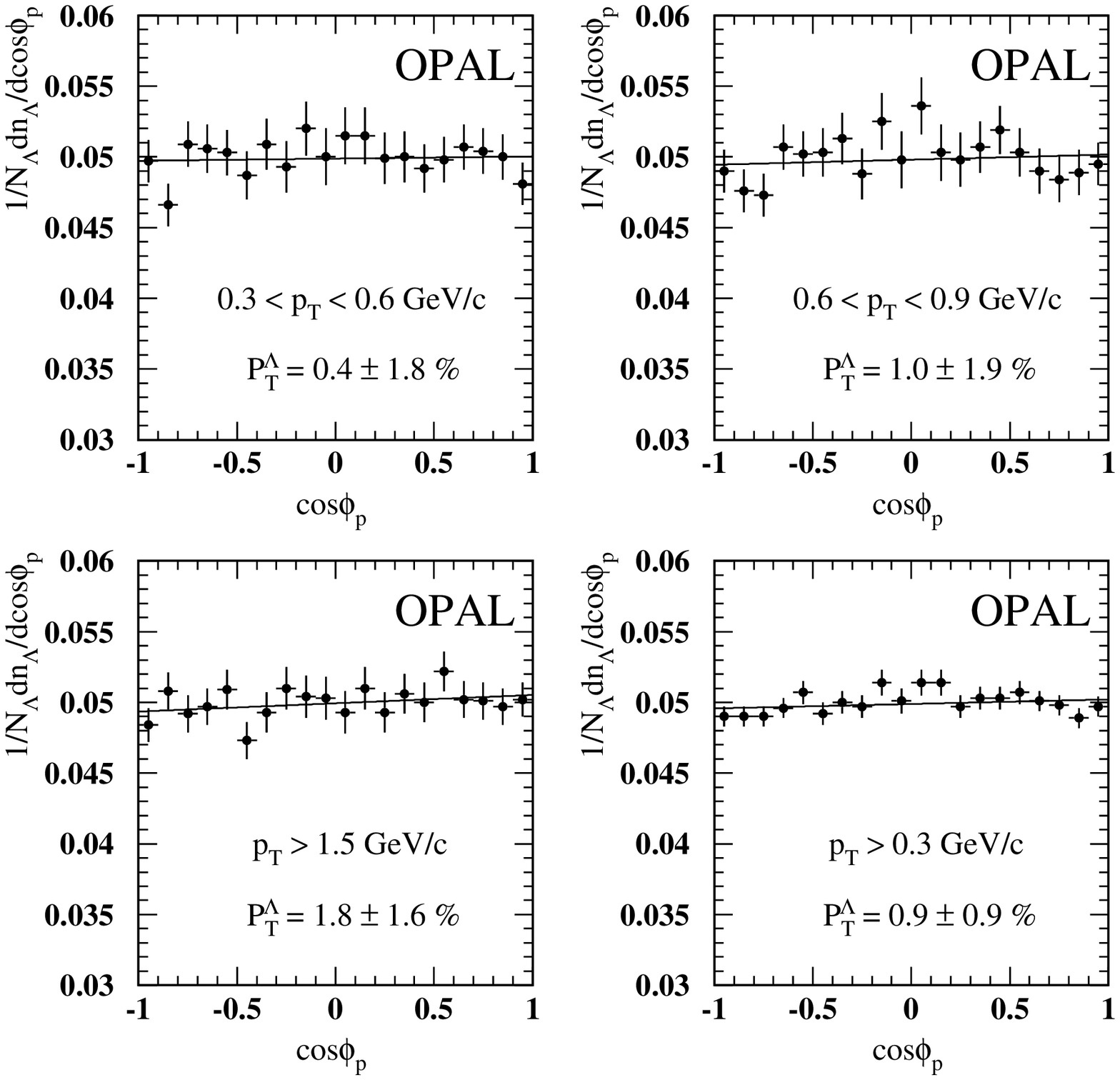,height=20cm,width=17.5cm}}
\end{center} 
\caption{The fits to the efficiency-corrected $\cos\phi_{p}$ distributions for
several $p_T$ regions. The error bars are statistical errors only.} 
\end{figure} 

\newpage 
\begin{figure}[htbp]
\begin{center} 
\mbox{\epsfig{file=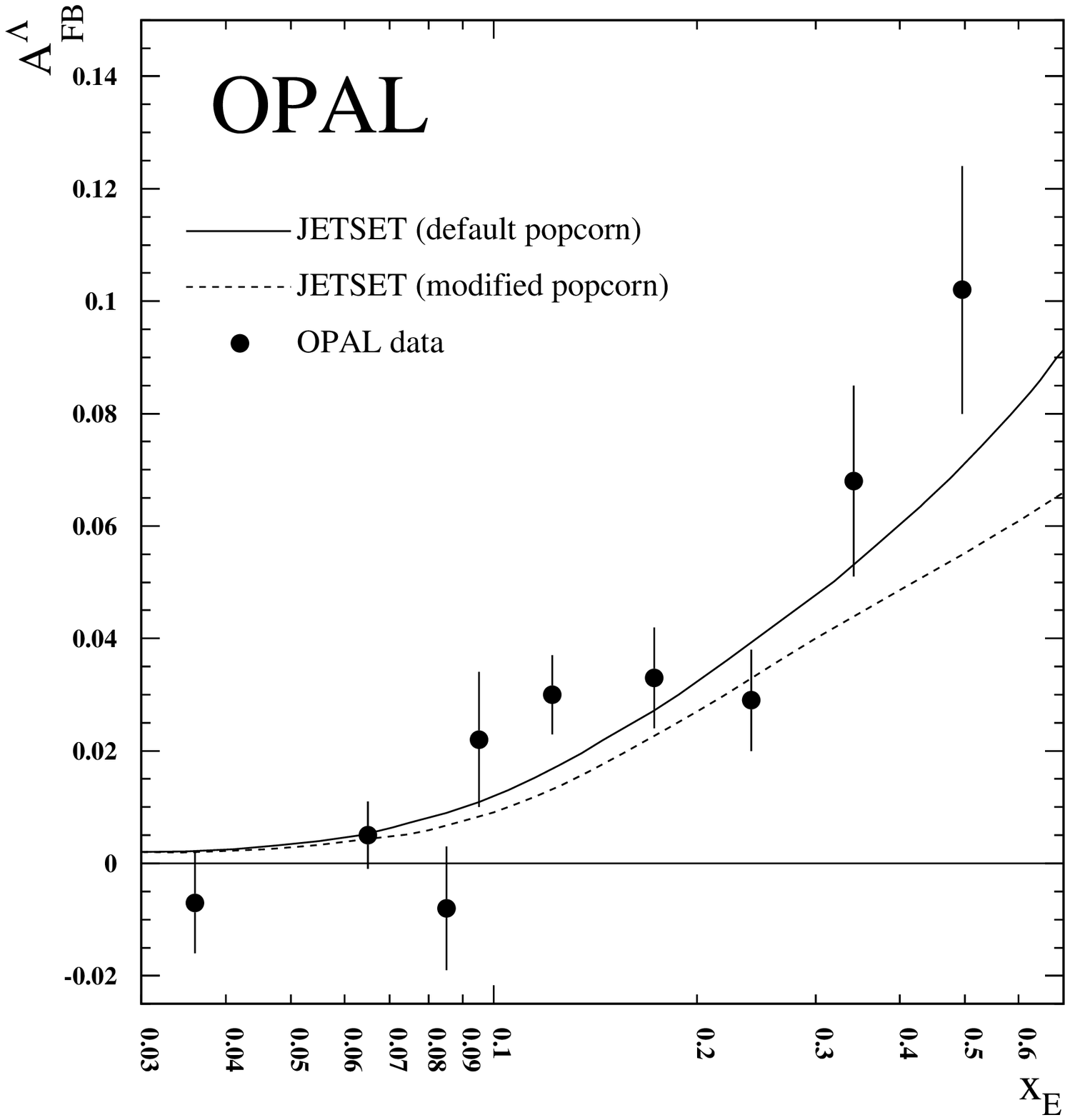,height=20cm,width=17.5cm}}
\end{center} 
\caption{The measured $\Lambda$ forward-backward asymmetry (solid points) and 
predictions from \JETSET\ with two different versions of the popcorn model of
baryon production. The error bars are statistical and systematic errors, added 
in quadrature.} 
\end{figure} 


\newpage 
\begin{thebibliography}{99}

\bibitem{lund_pred} 
G. Gustafson and J. H\"akkinen, Phys. Lett. {\bf B303} (1993) 350. 

\bibitem{gluon_rad} 
J.G. K\"orner, A. Pilaftsis and M.M. Tung, Z. Phys. {\bf C63} (1994) 575. 

\bibitem{renard} 
J.E. Augustin and F.M. Renard, {\em Proc. LEP Summer Study}, Report CERN 79-01,
Vol. 1 (1979) 185; \\ 
J.E. Augustin and F.M. Renard, Nucl. Phys {\bf B162} (1980) 341.
 
\bibitem{jetset}
T.~Sj\"ostrand, Comp. Phys. Comm. {\bf 39} (1986) 347; \\
T.~Sj\"ostrand and M.Bengtsson, Comp. Phys. Comm. {\bf 43} (1987) 367; \\
T.~Sj\"ostrand, Comp. Phys. Comm. {\bf 82} (1994) 74. 

\bibitem{PDG}
R.M. Barnett et al., Particle Data Group, Phys. Rev. {\bf D54} (1996) 1.

\bibitem{trans1} 
A.D. Panagiotou, Int. J. of Mod. Phys. {\bf A5} (1990) 1197. 

\bibitem{trans2} 
W. Lu, Phys. Rev. {\bf D51} (1995) 5305. 

\bibitem{light_quarks} 
OPAL Collaboration, K. Ackerstaff {\it et al.,} {\em  
Measurement of the Branching Fractions and Forward-Backward Asymmetries of 
the $\Zzero$ into Light Quarks}, CERN-PPE/97-063, to be published in Z. Phys. 
C.

\bibitem{opal}
OPAL Collaboration, M. Ahmet {\it et al.,} 
Nucl. Instr. and Meth. {\bf A305} (1991) 275.

\bibitem{dEdx}
M.~Hauschild {\it et al.,} Nucl. Instr. and Meth. {\bf A314} (1992) 74.

\bibitem{hadsel}
OPAL Collaboration, G.~Alexander {\it et al.,} Z. Phys. {\bf C52} (1991) 175.

\bibitem{gopal}
J. Allison {\it et al.,} Nucl. Instr. and Meth. {\bf A317} (1992) 47.

\bibitem{mctune}
OPAL Collaboration, P.D.~Acton {\it et al.,} Z. Phys. {\bf C58} (1993) 387; \\
OPAL Collaboration, R.~Akers {\it et al.,} Z. Phys. {\bf C69} (1996) 543.

\bibitem{baryons}
OPAL Collaboration, G.~Alexander {\it et al.,} Z. Phys. {\bf C73} (1997) 569.  

\bibitem{kzero} 
OPAL Collaboration, G.~Alexander {\it et al.,} Z. Phys. {\bf C67} (1995) 389.

\bibitem{popcorn} 
B. Andersson {\it et al.,} Physica Scripta {\bf 32} (1985) 574. 

\bibitem{new_popcorn} 
P. Ed\'en and G. Gustafson, Z. Phys. {\bf C75} (1997) 41; \\ 
P. Ed\'en, {\em A Program For Baryon Generation and Its Applications to Baryon
Fragmentation in DIS}, Lund Preprint LU TP 96-29. 

\bibitem{lambdab_pol} 
ALEPH Collaboration, D. Buskulic {\it et al.,} Phys. Lett. {\bf B365} (1996)
437. 

\bibitem{ALEPH} 
ALEPH Collaboration, D.~Buskulic {\it et al.,} Phys. Lett. {\bf B374} (1996)
319. 

\bibitem{lampol_th}
A. Kotzinian, A. Bravar and D. von Harrach, {\em $\Lambda$ and $\lbar$ 
Polarization in Lepton Induced Processes}, preprint hep-ph 9701384, to be 
published in Z. Phys. C.  

\bibitem{DELPHI} 
DELPHI Collaboration, P.~Abreu {\it et al.,} Z. Phys. {\bf C67} (1995) 1. 

\end{thebibliography}
\end{document}